\documentclass[]{interact}

\usepackage{epstopdf}
\usepackage[caption=false]{subfig}

\usepackage[numbers,sort&compress]{natbib}
\bibpunct[, ]{[}{]}{,}{n}{,}{,}
\makeatletter
\def\NAT@def@citea{\def\@citea{\NAT@separator}}
\makeatother

\theoremstyle{plain}

\theoremstyle{definition}

\theoremstyle{remark}

\begin{document}

\articletype{ARTICLE TEMPLATE}

\title{Frank Elastic Constants of Associating Rigid-Rodlike Systems
}
\author{
\name{Akihiko Matsuyama\thanks{CONTACT A.~Matsuyama. Email: aki\_mcyama@mac.com} }
\affil{Department of Physics and Information Technology, Faculty of Computer Science and Systems Engineering, 
Kyushu Institute of Technology, 
Kawazu 680-4, Iizuka, Fukuoka 820-8502, JAPAN}
}
\maketitle

\begin{abstract}
Frank elastic constants of associating rigid rodlike systems are investigated using a mean field theory combined with the Straley elastic free energy formalism. Incorporating the aggregation equilibrium and size-dependent orientational ordering, we analyze the concentration and temperature dependences of the elastic constants for both a monomer--dimer model and a fully polydisperse rodlike aggregate model. We find that aggregation selectively enhances the bend elastic constant through the long-tailed volume fraction distribution, resulting in pronounced elastic anisotropy, $K_{33}\gg K_{11}>K_{22}$. Decreasing temperature or increasing concentration promotes the formation of long rodlike aggregates and induces a rapid increase in the bend elasticity, while the relation $K_{11}=3K_{22}$ remains unchanged.

\end{abstract}


\section{Introduction}
Associating rodlike molecules and self-assembled aggregates exhibit a wide variety of liquid-crystalline phases and play important roles in soft matter systems~\cite{Mezzenga05,kato02,Rey2019,Rey10,Matsu26crc}. Typical examples include chromonic liquid crystals~\cite{Lydon,Horowitz05,Tortoraa11,Tomasik}, supramolecular DNA polymers~\cite{Yevdokimov,Nakata07,Michele16}, micellar and lyotropic nematic systems~\cite{cates90,Bodenn94,Radlev78,Kuntz08,Johannesson96,Char79,Hendrikx81,Quist92}, and self-assembled nanotubular aggregates~\cite{Ciferri99,Lee09,Bong01,Shimizu05}. In these systems, the aggregation equilibrium strongly affects the orientational ordering, rheological behavior, and elastic properties of the liquid-crystalline phases. Understanding how aggregation influences the Frank elastic constants is therefore an important problem in the theory and experiment of self-assembling liquid crystals.

The orientational elasticity of nematic liquid crystals is commonly described by the Frank elastic free energy, which consists of the splay, twist, and bend elastic constants $K_{11}$, $K_{22}$, and $K_{33}$. For nonassociating rigid rod systems, 
Straley developed a microscopic theory of the Frank
elastic constants based on excluded-volume interactions
between anisotropic particles~\cite{Onsager,Straley73,Lee85}. The Straley theory shows the characteristic elastic hierarchy
$
K_{33}>K_{11}>K_{22},
$
as well as the relation
$
K_{11}/K_{22}=3.
$ 
Later, Odijk derived the analytical expressions for the elastic constants and showed that, in the strong-alignment limit, the bend elastic constant exhibits a stronger concentration dependence than the splay and twist elastic constants~\cite{Odijk86}.

The Frank elastic constants of nonassociating rod systems
have been investigated theoretically in considerable detail~\cite{Straley73,Lee85,Odijk86,Priest73,Ferrarini10,Poniewierski,Sato005}.
In contrast, much less is understood about how aggregation
affects the elastic properties of associating liquid-crystalline
systems. In these systems, the volume fraction distribution of aggregates generally extends to large aggregate sizes, and the orientational ordering strongly depends on the aggregation equilibrium~\cite{Herzfeld81,Gelbart85,Odijk87,McMullen85,Herzfeld88,Taylor90,Taylor93,Schoot94,Lu06,Michele19,Michele12,Michele12a}. Since the excluded volume between rodlike particles scales approximately with the product of their lengths~\cite{Book_de}, long aggregates are expected to exert a strong orientational influence on surrounding shorter aggregates. Such collective excluded-volume interactions may strongly modify the elastic anisotropy of the nematic phase. Experimental studies of chromonic liquid crystals have reported distinctive elastic properties and pronounced elastic anisotropy~\cite{Zhou14,Zhou12}. Related elastic properties have also been investigated in polymer nematic liquid crystals~\cite{Taratuta88,Vroege88}. These observations suggest that aggregation plays an essential role in controlling orientational elasticity.

Previous theoretical studies have investigated
associating liquid-crystalline systems, focusing on
different aspects of aggregate formation and flexibility. De Michele et al.~\cite{Michele16} investigated the cholesteric ordering arising from the association of DNA oligomers, focusing on the propagation of chirality. Romani et al.~\cite{Ferrarini18} studied chromonic liquid crystals in which molecules self-assemble into semiflexible chains, explicitly accounting for the intrinsic flexibility of the aggregates. Wensink~\cite{Wensink19} developed a theory of associating rods that accounts for molecular flexibility and chirality. In contrast to these approaches, the present work focuses on perfectly rigid rodlike aggregates and applies Straley's theory of Frank elastic constants to associating systems, thereby providing a complementary description in the rigid-aggregate limit.

Recently, we have presented a mean field theory to describe the nematic-isotropic phase transition (NIT) of associating rodlike aggregates~\cite{Matsu25}. By combining the Onsager theory for rodlike particles~\cite{Onsager} with the Flory--Huggins theory for polydisperse polymer solutions~\cite{Flory53,tanakaf}, we derived the free energy of associating rodlike aggregates and calculated the mean aggregation number and orientational order parameters.

In this paper, we theoretically study the Frank elastic constants of associating rigid rodlike systems using the previously developed mean field theory~\cite{Matsu25} combined with the Straley elastic formalism~\cite{Straley73}. Both a monomer--dimer model and a fully polydisperse rodlike aggregate model are considered. The volume fraction distribution of aggregates and orientational ordering, as well as the concentration and temperature dependences of the elastic constants, are systematically analyzed. Particular attention is paid to the role of aggregation in modifying the elastic anisotropy of the nematic phase.
Our results show that aggregation selectively enhances the bend elastic constant and produces an elastic anisotropy, $K_{33}\gg K_{11}>K_{22}$. We demonstrate that this giant bend elasticity originates from the long-tail structure of the volume fraction distribution, in which a small population of long aggregates dominates the collective orientational response through excluded-volume interactions. The present theory therefore provides a microscopic mechanism by which aggregation controls the elastic properties of nematic liquid crystals.

The remainder of this paper is organized as follows. In Section~\ref{sec2}, we introduce the Frank elastic constants of self-assembled rodlike aggregates. In Section~\ref{sec3}, we present numerical results for the elastic constants, considering three cases: nonassociating rigid rodlike molecules in Sec.~\ref{sub1}, the monomer--dimer model in Sec.~\ref{sub2}, and the polydisperse rodlike aggregate model in Sec.~\ref{sub3}. Finally, Section~\ref{sec4} summarizes our findings.

\section{Elastic constants of self-assembled rodlike aggregates}
\label{sec2}

\begin{figure}
\begin{center}
\includegraphics*[height=8cm]{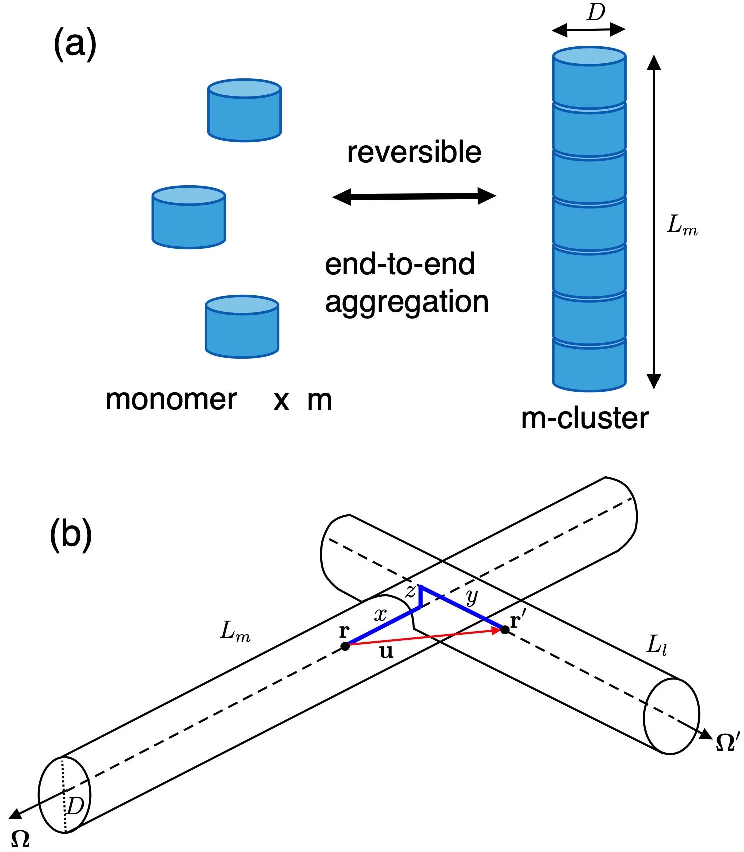}
\caption{ (Colour online) (a) Schematic representation of rodlike $m$-clusters formed by the end-to-end aggregation of $m$ monomer molecules. The system consists of polydisperse, anisotropic, rodlike aggregates that can exhibit liquid-crystalline ordering.
(b) Geometry used for calculating the excluded volume interaction between $m$- and $l$-clusters with lengths $L_m$ and $L_l$, respectively. The relative coordinates are defined by $-L_m/2<x<L_m/2$, $-L_l/2<y<L_l/2$, and $-D<z<D$.
}
\label{fig1}
\end{center}
\end{figure}

We consider a binary mixture of self-assembled solute molecules and solvent molecules~\cite{Matsu25}. Each solute molecule can reversibly form rigid rodlike aggregates consisting of $m$ solute molecules, referred to as $m$-clusters, as shown in Fig. \ref{fig1}(a).
Let $N_m$ be the number of $ m$-clusters ($ m=1,2,\cdots, \infty$) and $N_0$ be the number of solvent molecules. 
Let $L_m$ and $D$ denote the length and diameter of an $m$-cluster, respectively. The corresponding axial ratio is given by $L_m/D=nm$, and $n=(L/D)$ is the axial ratio of the monomer molecule. 
We define the reference volume $v_0=(\pi/4)D^3$. The volume of an $m$-cluster is then given by $v_m=(\pi/4)D^2L_m=v_0nm$. We assume that the volume of the solvent molecule is given by $v_s=v_0$. The total volume of our system is given by $V=v_0N_t$, where 
\begin{equation}
N_t=N_0+n\sum_{m=1}^{\infty} mN_{m},
 \label{eq200}
\end{equation}
is the total number of lattice cells in the system. The volume fraction of the $m$-cluster is given by
\begin{equation}
\phi_m=v_0nm N_m/V=nmN_m/N_t,
 \label{eq201}
\end{equation}
and that of the solvent is 
\begin{equation}
\phi_0=v_0 N_0/V=N_0/N_t.
 \label{eq202}
\end{equation}
The total volume fraction $\phi$ of the solute molecules, or $m$-clusters, is given by
\begin{equation}
\phi=\sum_{m=1}^{\infty} \phi_{m}=1-\phi_0.
 \label{eq203}
\end{equation}
The mean aggregation number is defined as
 \begin{equation}
\langle m \rangle=\sum_{m=1}^{\infty}m N_{m}/\sum_{m=1}^{\infty} N_{m}=\frac{\phi}{n}/\sum_{m=1}^{\infty}\frac{\phi_m}{nm}.
 \label{eq217a}
\end{equation}

Let $f_m({\bf \Omega}\cdot {\bf n}({\bf r})))$ be the orientational distribution function of the $m$-cluster, where ${\bf n}$ is the director and ${\bf\Omega}$ is the orientation vector of the $m$-cluster at a position ${\bf r}$ (see Fig. \ref{fig1} (b)).
Taking into account the excluded volume interactions~\cite{McMullen85,Schoot94,Onsager} between rodlike aggregates for the nematic ordering, the nematic free energy $F_{nem}$ of associating rodlike molecules is given by~\cite{Matsu25}
\begin{eqnarray}
\beta F_{nem}&= &\sum_{m=1}^{\infty} \rho_m\iint f_m({\bf \Omega}\cdot {\bf n}({\bf r}))\ln [ 4\pi f_m({\bf \Omega}\cdot {\bf n}({\bf r}))] d\Omega d{\bf r}
\nonumber \\
&&
+\frac{1}{2} \sum_{m=1}^{\infty} \sum_{l=1}^{\infty} \rho_m \rho_l \iiiint f_m({\bf \Omega}\cdot {\bf n}({\bf r})) f_l({\bf \Omega}'\cdot {\bf n}({\bf r}'))
\nonumber \\
&&
 \times\delta_{ml}({\bf r}, {\bf r}'; {\bf \Omega}, {\bf \Omega}')d\Omega d\Omega'd{\bf r}d{\bf r}',
 \label{eq207}
\end{eqnarray}
where \(\beta=1/(k_{\mathrm B}T)\), \(k_{\mathrm B}\) is the Boltzmann constant, and \(T\) is the absolute temperature; \(\rho_m=\phi_m/(v_0nm)\) is the number density of the \(m\)-cluster, as defined in Eq.~(\ref{eq201}), and $d\Omega=\sin\theta d\theta d\psi$ is the solid angle, where $0<\theta<\pi$ and $0<\psi<2\pi$. The first term in Eq.~(\ref{eq207}) shows the entropy changes due to the orientational ordering of the $m$-cluster. 
The last term $\delta_{ml}(\cdots)\propto |{\bf \Omega}\times {\bf\Omega}'|$ shows the excluded volume kernel between the rodlike aggregate of length $L_m$ and that of length $L_l$ \cite{Odijk87,McMullen85}. 

The extra free energy due to variation of the director ${\bf n}({\bf r})$ is given by
\begin{eqnarray}
\beta \Delta F &= &
\frac{1}{2} \sum_{m=1}^{\infty} \sum_{l=1}^{\infty} \rho_m \rho_l \iiiint f_m({\bf \Omega}\cdot {\bf n}({\bf r})) \delta _{ml}({\bf r}, {\bf r}'; {\bf\Omega}, {\bf\Omega}')
 \nonumber \\
 &&
\times \big[ f_l({\bf \Omega}'\cdot {\bf n}({\bf r}')) - f_l({\bf \Omega}'\cdot {\bf n}({\bf r}))\big]d{\bf r} d{\bf r}'d\Omega d\Omega'.
 \label{eq301}
\end{eqnarray}
We will now calculate Eq. (\ref{eq301}) to second order in the derivatives of ${\bf n}({\bf r})$:
\begin{eqnarray}
f_l({\bf \Omega}'\cdot {\bf n}({\bf r}'))-f_l({\bf \Omega}'\cdot {\bf n}({\bf r}))&=&
f_l'({\bf \Omega}'\cdot {\bf n}({\bf r}))\big[ ({\bf u} \cdot \nabla)({\bf \Omega}'\cdot {\bf n}({\bf r}))+\frac{1}{2} ({\bf u} \cdot \nabla)^2({\bf \Omega}'\cdot {\bf n}({\bf r}))\big]
\nonumber \\
&&
+\frac{1}{2} f_l''({\bf \Omega}'\cdot {\bf n}({\bf r}))\big[ ({\bf u} \cdot \nabla)({\bf \Omega}'\cdot {\bf n}({\bf r}))\big]^2+\cdots,
 \label{eq302}
\end{eqnarray}
where ${\bf u}={\bf r}'-{\bf r}$, $f_l'$, and $f_l''$ show the first and second derivatives of the distribution function with respect to its argument.
Introducing $\eta={\bf \Omega}'\cdot\mathbf n,$
the two second-order terms in Eq.~(\ref{eq302}) can be combined by using the equation
\begin{equation}
\partial_i
\left[
f_l'(\eta)\,
\partial_j\eta
\right]
=
f_l''(\eta)\,
(\partial_i\eta)
(\partial_j\eta)
+
f_l'(\eta)\,
\partial_i\partial_j\eta,
\label{identity}
\end{equation}
which follows directly from the product and chain rules.
Substituting this expression into the free energy  Eq. (\ref{eq301}) and performing integration by parts, while noting that the contact function
$\delta_{ml}(\mathbf{u}; {\bf\Omega},{\bf\Omega}')$
and the relative displacement $\mathbf{u}$ are independent of the spatial coordinate $\mathbf r$, we obtain
\begin{eqnarray}
\beta \Delta F &= &
\frac{1}{2} \sum_{m=1}^{\infty} \sum_{l=1}^{\infty} \rho_m \rho_l \iiiint f_m({\bf \Omega}\cdot {\bf n}({\bf r})) \delta _{ml}({\bf u}; {\bf\Omega}, {\bf \Omega}')f_l'({\bf \Omega}'\cdot {\bf n}({\bf r}))
 \nonumber \\
 &&
\times \big[ ({\bf u}\cdot \nabla)({\bf \Omega}'\cdot {\bf n}({\bf r})) \big] d{\bf r} d{\bf u}d\Omega d\Omega'
 \nonumber \\
 &&
-\frac{1}{4}\sum_{m=1}^{\infty} \sum_{l=1}^{\infty} \rho_m \rho_l  \iiiint \big[ ({\bf u}\cdot \nabla)({\bf \Omega}\cdot {\bf n}({\bf r})) \big]f_m'({\bf \Omega}\cdot {\bf n}({\bf r}))
 \nonumber \\
 &&
\times \delta _{ml}({\bf u}; {\bf \Omega}, {\bf \Omega}')f_l'({\bf \Omega'}\cdot {\bf n}({\bf r}))\big[ ({\bf u}\cdot \nabla)({\bf \Omega'}\cdot {\bf n}({\bf r})) \big]d{\bf r} d{\bf u} d\Omega d\Omega',
 \label{eq303}
\end{eqnarray}
where integration by parts has been used to combine the second-order terms.
The first term can be neglected for the uniaxial nematic phase and will automatically vanish if $\delta_{ml}(\cdots)$
is an even function of ${\bf u}$; this corresponds to the case of nematic liquid crystals. 
Thus, the second-order free-energy contribution is expressed solely in terms of the first derivatives of the orientational distribution functions and the director field, providing the starting point for the evaluation of the Frank elastic constants.
The coefficient $1/4$ appearing in Eq.~(\ref{eq303}) differs from that given by Straley~\cite{Straley73}.
Our derivation follows from the identity Eq.~(\ref{identity}) and an explicit integration by parts.
The resulting prefactor agrees with the corrected expression employed by Lee and Meyer~\cite{Lee85},
who noted a missing factor of 1/2 in the previous paper.

The excluded volume interaction is given by
 \begin{eqnarray}
\delta _{ml}({\bf u}; {\bf \Omega}, {\bf \Omega}')=|{\bf \Omega}\times {\bf \Omega}'|  W({\bf u}),
\label{A6}
\end{eqnarray}
 where $W({\bf u})$ is a window function and is given by:
\begin{equation}
W(\mathbf{u})=
\begin{cases}
1, & |x|<L_m/2,\quad |y|<L_l/2,\quad |z|<D,\\[2mm]
0, & \text{otherwise}.
\end{cases}
\label{A7}
\end{equation}
By integrating Eq. (\ref{A6}) over the relative position vector with $d{\bf u}=dxdydz$, we obtain
\begin{equation}
\int d\mathbf{u}\,\delta_{ml}
=2L_mL_lD|\mathbf{\Omega}\times\mathbf{\Omega}'|,
\label{excluded}
\end{equation}
which reproduces the leading-order excluded volume of two long rigid rods of lengths \(L_m\) and \(L_l\) and diameter \(D\) \cite{Schoot94}. In the special case of \(L_m=L_l=L\), Eq. (\ref{excluded}) reduces to Straley's excluded-volume expression for two identical rods.

The distance vector ${\bf u}$ between the two rods, as shown in Fig.~\ref{fig1}(b), is given by
\begin{eqnarray}
{\bf u}&= &x {\bf \Omega}+y {\bf \Omega}'+z({\bf \Omega}\times {\bf \Omega}')|{\bf \Omega}\times {\bf \Omega}'|^{-1},
 \label{eq304}
\end{eqnarray}
Since $x$, $y$, $z$ are independent variables, using Eq. (\ref{eq304}), the integration is given by 
\begin{eqnarray}
\int \delta_{ml}({\bf u}; {\bf \Omega}, {\bf \Omega}')({\bf u}\cdot {\bf a}) ({\bf u}\cdot {\bf b})d{\bf u}&=&
\frac{1}{6}D |{\bf \Omega}\times {\bf \Omega}'|\big[L_lL_m^3({\bf \Omega}\cdot {\bf a})({\bf \Omega}\cdot {\bf b}) 
\nonumber \\
 &&
+L_l^3L_m({\bf \Omega}'\cdot {\bf a})({\bf \Omega}'\cdot {\bf b}) \big]
\nonumber \\
 &&
 +\frac{2}{3}L_lL_m D^3 \frac{\big[ ({\bf \Omega}\times{\bf \Omega}')\cdot {\bf a}\big] \big[ ({\bf \Omega}\times{\bf \Omega}')\cdot {\bf b}\big]}{ |{\bf \Omega}\times {\bf \Omega}'|},
 \label{eq305} 
\end{eqnarray}
for any vectors ${\bf a}$ and  ${\bf b}$. We here neglect the last term for $D/L_m\ll 1$~\cite{Straley73}, thus, the integration of Eq.~(\ref{eq303}) gives
\begin{eqnarray}
\beta \Delta F &= &
-\frac{D}{24}\sum_{m=1}^{\infty} \sum_{l=1}^{\infty} \rho_m \rho_l  \iiint f_m'({\bf \Omega}\cdot {\bf n}({\bf r})) f_l'({\bf \Omega'}\cdot {\bf n}({\bf r}))|{\bf \Omega}\times {\bf \Omega}'|
 \nonumber \\
 &&
 \times G_i(m,l) d{\bf r} d\Omega d\Omega',
 \label{eq306}
\end{eqnarray}
and 
\begin{eqnarray}
G_i(m,l)  &= &
L_m^3L_l \big[ ({\bf \Omega}\cdot \nabla) ({\bf \Omega}\cdot {\bf n})\big]\big[ ({\bf \Omega}\cdot \nabla) ({\bf \Omega}' \cdot {\bf n})\big]
 \nonumber \\
 &&
+L_mL_l^3 \big[ ({\bf \Omega}'\cdot \nabla) ({\bf \Omega}\cdot {\bf n})\big]\big[ ({\bf \Omega}'\cdot \nabla) ({\bf \Omega}' \cdot {\bf n})\big],
 \label{eq307}
\end{eqnarray}
where the explicit mention of the ${\bf r}$ dependence of ${\bf n}$ has been dropped and the $G_i(m,l)$ are geometrical factors corresponding to splay ($i=1$), twist ($i=2$), and bend ($i=3$) deformations, respectively. When $l=m=1$, Eq.~(\ref{eq306}) results in the Straley model~\cite{Straley73}.
Note that this result involves only the gradients of ${\bf n}$: $({\bf \Omega}\cdot \nabla) ({\bf \Omega}\cdot {\bf n})=\Omega_{\alpha}\Omega_{\beta}(\partial_{\alpha} n_{\beta})$.

Thus, Eq.~(\ref{eq306}) is already of second order in the director distortions and can be directly compared with the Frank elastic free energy:
\begin{eqnarray}
 \Delta F &= &
\frac{1}{2}\int \big[ K_{11} (\nabla\cdot {\bf n})^2+K_{22} ({\bf n}\cdot (\nabla\times  {\bf n}))^2
+K_{33} ({\bf n}\times  (\nabla\times  {\bf n}))^2\big] d{\bf r}.
 \label{eq308}
\end{eqnarray}
By comparing Eq.~(\ref{eq306})  with the Frank elastic free energy Eq.~(\ref{eq308}), the elastic constants can be expressed as
\begin{eqnarray}
 K_{ii}&= &
-\frac{c_1}{12}\sum_{m=1}^{\infty} \sum_{l=1}^{\infty} \phi_m \phi_l  \iint f_m'({\bf \Omega}\cdot {\bf n}({\bf r})) f_l'({\bf \Omega'}\cdot {\bf n}({\bf r}))|{\bf \Omega}\times {\bf \Omega}'|
 \nonumber \\
 &&
 \times {\tilde G}_i(m,l)  d\Omega d\Omega',
 \label{eq309}
\end{eqnarray}
\begin{eqnarray}
{\tilde G}_i(m, l)  &= &
m^2\big[ ({\bf \Omega}\cdot \nabla) ({\bf \Omega}\cdot {\bf n})\big]\big[ ({\bf \Omega}\cdot \nabla) ({\bf \Omega}' \cdot {\bf n})\big]
 \nonumber \\
 &&
+l^2 \big[ ({\bf \Omega}'\cdot \nabla) ({\bf \Omega}\cdot {\bf n})\big]\big[ ({\bf \Omega}'\cdot \nabla) ({\bf \Omega}' \cdot {\bf n})\big],
 \label{eq3010}
\end{eqnarray}
where $c_1=k_BT{n^2}/((\pi/4)^2D)$ and we have expressed it as a function of the volume fraction $\phi_m(=\pi(D/2)^2L_m\rho_m)$ and $\phi_l(=\pi(D/2)^2L_l\rho_l)$.
We define the reduced elastic constants as ${\tilde K}_{ii} =K_{ii}/c_1$.
Note that the geometrical factor ${\tilde G}_i(m, l) $ depends on the aggregation number $m$ and $l$.

It is worth noting that the general expression for the Frank elastic constants obtained in the present mean-field theory has a mathematical structure similar to that derived by Poniewierski and Stecki using density functional theory~\cite{Poniewierski}. In their theory, the Frank elastic constants are expressed in terms of the direct correlation function and the orientational single-particle distribution function. In both approaches, the elastic constants involve products of the first derivatives of two orientational distribution functions, as seen in Eq.~(\ref{eq309}). This similarity reflects the common feature of deriving the elastic constants from the second-order spatial variations of the free energy, although the underlying theoretical frameworks and interaction kernels are different. In the present theory, the interaction is described by a Straley-type excluded-volume interaction, and the volume fraction distribution is explicitly incorporated through the double summation over the aggregation numbers $m$ and $l$. This formulation allows us to investigate the effects of aggregate polydispersity on the Frank elastic constants.

We here take that the director ${\bf n}$ is parallel to the $x$-axis:
\begin{eqnarray}
{\bf \Omega}= (\cos\theta, \sin\theta\cos\varphi, \sin\theta\sin\varphi),
 \label{eq301a}
\end{eqnarray}
where $\theta$ is the polar angle measured from the director ${\bf n}$, and $\varphi$ is the azimuthal angle around the $x$-axis. Using Eq. (\ref{eq301a}), the excluded volume kernel is given by
\begin{eqnarray}
|{\bf \Omega}\times {\bf \Omega}'| =| \sin\gamma_{12}|=\sqrt{1-({\bf \Omega}\cdot {\bf \Omega}')^2},
 \label{eq_sin}
\end{eqnarray}
where $\gamma_{12}$ is the angle between the two unit vectors ${\bf \Omega}$ and ${\bf \Omega}'$.

Retaining only terms linear in the distortion coordinates, for the splay $(i=1)$ deformation, we take ${\bf n}=(1, y, 0)$ and the geometrical factor is given by
\begin{eqnarray}
{\tilde G}_1(m,l) = 
({\bf \Omega}\cdot {\bf e}_y) ({\bf \Omega}'\cdot {\bf e}_y)
\big[ m^2 ({\bf \Omega}\cdot {\bf e}_y)^2+l^2 ({\bf \Omega}' \cdot {\bf e}_y)^2\big],
 \label{eq311}
\end{eqnarray}
for the twist $(i=2)$: ${\bf n}=(1, z, 0)$,
\begin{eqnarray}
{\tilde G}_2(m,l) = 
({\bf \Omega}\cdot {\bf e}_y) ({\bf \Omega}'\cdot {\bf e}_y)
\big[ m^2 ({\bf \Omega}\cdot {\bf e}_z)^2+l^2 ({\bf \Omega}' \cdot {\bf e}_z)^2\big],
 \label{eq312}
\end{eqnarray}
and for the bend $(i=3)$: ${\bf n}=(1, x, 0)$,
\begin{eqnarray}
{\tilde G}_3(m,l) = 
({\bf \Omega}\cdot {\bf e}_y) ({\bf \Omega}'\cdot {\bf e}_y)
\big[ m^2 ({\bf \Omega}\cdot {\bf e}_x)^2+l^2 ({\bf \Omega}' \cdot {\bf e}_x)^2\big],
 \label{eq313}
\end{eqnarray}
where ${\bf e}_{\alpha}$ ($\alpha=x,y,z$) denotes the unit vectors of the $x$-, $y$-, and $z$-axes. When $l=m=1$, Eqs.~(\ref{eq311})--(\ref{eq313}) result in the Straley model~\cite{Straley73}.

The geometrical factors in Eqs.~(\ref{eq311})--(\ref{eq313}) provide
insight into the origin of the enhancement of the elastic constants
in associating systems.
In particular, these factors contain terms
proportional to $m^2$ and $l^2$, which reflect the
lengths of the interacting aggregates. Thus, long
aggregates can make large contributions to the elastic constants, even when
they interact with shorter aggregates.
This effect is particularly important for the bend
deformation. Since the director is oriented along
the $x$-axis, the terms in the square brackets of
the geometrical factor for bend involve the longitudinal
components $(\mathbf{\Omega}\cdot\mathbf e_x)^2$ and
$(\mathbf{\Omega}'\cdot\mathbf e_x)^2$,
whereas those for splay and twist involve the
corresponding transverse components.
In the strongly aligned state, the longitudinal components
are large: $(\Omega_x)^2\simeq 1$, while the transverse components are
relatively small: $(\Omega_y)^2, (\Omega_z)^2 \ll 1$. Then, the length-dependent
terms in the bend geometrical factor can produce strong contributions from long
aggregates, leading to an enhancement of $K_{33}$.

\subsection{Distribution functions of $m$-cluster}
In the numerical calculations, we employ the orientational distribution function of the $m$-cluster, derived from Eq.~(\ref{eq207}) in our previous mean field theory for associating rodlike aggregates~\cite{Matsu25,Matsu26crc}:
\begin{equation}
f_m(\eta)=\frac{1}{Z_m}\exp\bigg( \frac{5}{4}nm\phi \langle S \rangle  P_2(\eta)  \bigg),
 \label{n2}
\end{equation}
where $P_2(\eta)=(3/2)\eta^2-1/2$ and $\eta={\bf \Omega}\cdot {\bf n}({\bf r})$  with the normalization constant $Z_m$, satisfying $\int f_m(\eta) d\Omega=1$. Note that $f'_m(\eta)=(15/4)nm\phi \langle S \rangle \eta f_m(\eta)$.
The average value $\langle S \rangle$ of the orientational order parameter $S_m$ of the $m$-cluster is given by
\begin{equation}
 \langle S \rangle= \sum_{m=1}^{\infty} S_m \phi_m/  \sum_{m=1}^{\infty}  \phi_m= \sum_{m=1}^{\infty} S_m \phi_m/  \phi.
 \label{n3}
\end{equation}
Using the orientational distribution function (Eq.~(\ref{n2})), the orientational order parameter of the $m$-cluster is given by
\begin{eqnarray}
S_m&=&\int P_2(\eta) f_m(\eta) d\Omega
 \nonumber \\
 &=&
I_{[1,m]}/I_{[0,m]},
 \label{n1}
\end{eqnarray}
where we define the function
\begin{equation}
I_{[q,m]}=\int_0^1\big[P_2(\eta)\big]^q\exp\bigg( \frac{5}{4}nm\phi \langle S \rangle  P_2(\eta)\bigg) d\eta, 
 \label{eq211ab}
\end{equation}
with $q=0, 1$ and the normarization constant in Eq.~(\ref{n2}) is given by $Z_m=4\pi I_{[0,m]}$.

Using the chemical equilibrium conditions for $m$-cluster formation~\cite{tanakaf}, the volume fraction of the $m$-cluster is given by~\cite{Matsu25}
\begin{equation}
 \phi_m=\frac{m}{K_a}(\phi_1K_a)^m ,
 \label{eq220c}
\end{equation}
where we define the association constant
 \begin{equation}
 K_a=\omega\exp(\beta\epsilon),
 \label{eq220dd}
\end{equation}
for a single bond formation with a constant $\omega$, related to the local entropy change: $s_0=k_B\ln \omega$, and $\epsilon (> 0)$ is the bonding energy associated with the formation of a single bond.

For given values of the total concentration $\phi$ and the association constant $K_a$ (or temperature), the volume fraction distribution $\phi_m$ is determined from Eqs.~(\ref{eq203}) and (\ref{eq220c}). The orientational order parameters $S_m$ and $\langle S\rangle$ are then determined self-consistently from Eqs.~(\ref{n2})--(\ref{eq211ab}).
Unlike in our previous work~\cite{Matsu25}, where the volume fraction of an $m$-cluster is proportional to a factor $D_m(S)$ that depends on the orientational order parameter $S$, we set $D_m(S)=1$ in the present model. This allows us to examine the effects of the volume fraction distribution on the Frank elastic constants without including the coupling between aggregation and orientational ordering.

\section{Numerical results and discussion}
\label{sec3}
In this section, we first examine the Straley model
for nonassociating rigid rodlike molecules to validate
the present numerical scheme (Sec.~\ref{sub1}).
We then consider a simple monomer--dimer model as
an introductory example of an associating system
(Sec.~\ref{sub2}), followed by a polydisperse model
of self-assembled rodlike aggregates (Sec.~\ref{sub3}).
The concentration and temperature dependences of
the Frank elastic constants are analyzed in detail.

\subsection{Nonassociating model of rigid rodlike molecules}
\label{sub1}
To validate the numerical procedure, we calculate the Frank elastic constants, using Eq.~(\ref{n2}), for a nonassociating rod system with $n=L/D=50$, $\phi=0.1$, $T=300$ K and $m=l=1$, corresponding to the conditions considered by Straley~\cite{Straley73}. The present calculation yielde $S=0.765$, $K_{11}^{non}=2.1~\mathrm{pN}$, $K_{22}^{non}=0.69$ $\mathrm{pN}$, and $K_{33}^{non}=9.7~\mathrm{pN}$, together with the characteristic ratio $K_{11}^{non}/K_{22}^{non}=3$. These values are approximately one-half of the corresponding values reported by Straley, while preserving the characteristic elastic hierarchy $K_{33}^{non}>K_{11}^{non}>K_{22}^{non}$. This difference is consistent with the missing factor of 1/2 pointed out by Lee and Meyer~\cite{Lee85}. These results confirm the validity of the present numerical implementation.

In the following, we set $n=L/D=20$ and consider the nonassociating monomeric state with $m=l=1$.
In this case, a nematic--isotropic transition (NIT) takes place at $\phi \simeq 0.18$. Above the NIT concentration, the orientational order parameter $S(=S_1)$ of the rods increases with increasing concentration $\phi$.

\begin{figure}
\begin{center}
\includegraphics*[height=5cm]{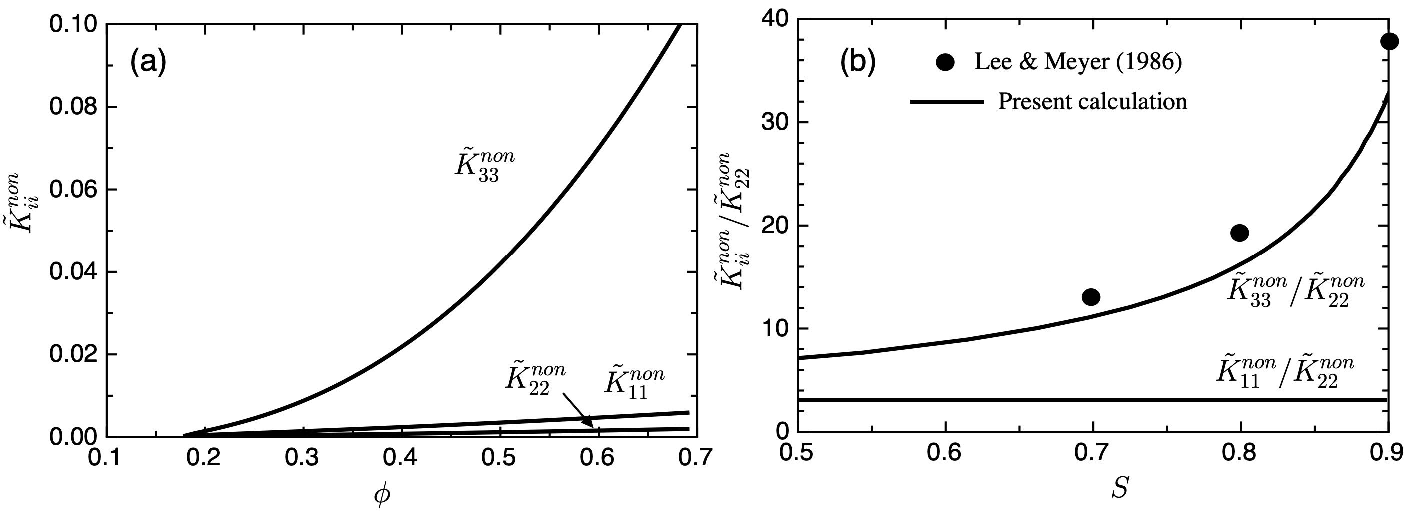}
\caption{
Frank elastic constants for the nonassociating rod system with an aspect ratio $L/D=20$. (a) Reduced elastic constants $\tilde{K}_{11}^{\mathrm{non}}$, $\tilde{K}_{22}^{\mathrm{non}}$, and $\tilde{K}_{33}^{\mathrm{non}}$, plotted against $\phi$. (b) Ratios $\tilde{K}_{33}^{\mathrm{non}}/\tilde{K}_{22}^{\mathrm{non}}$ and $\tilde{K}_{11}^{\mathrm{non}}/\tilde{K}_{22}^{\mathrm{non}}$, plotted against the orientational order parameter $S$. The solid curves represent the present calculations, while the filled circles denote the numerical results reported by Lee and Meyer~\cite{Lee85}.
}
\label{fig2}
\end{center}
\end{figure}

Figure 2(a) shows the reduced elastic constants $\tilde{K}_{11}^{\mathrm{non}}$, $\tilde{K}_{22}^{\mathrm{non}}$, and $\tilde{K}_{33}^{\mathrm{non}}$, plotted against $\phi$. 
As the concentration increases, all three elastic constants increase monotonically. However, the bend elastic constant $\tilde{K}_{33}^{\mathrm{non}}$ grows much more rapidly than the splay and twist elastic constants, whereas $\tilde{K}_{11}^{\mathrm{non}}$ and $\tilde{K}_{22}^{\mathrm{non}}$ remain comparatively small over the entire range. This behavior reflects the strong elastic anisotropy characteristic of highly ordered rodlike nematic systems.
To examine the elastic anisotropy, Figure 2(b) shows the ratios $\tilde{K}_{33}^{\mathrm{non}}/\tilde{K}_{22}^{\mathrm{non}}$ and $\tilde{K}_{11}^{\mathrm{non}}/\tilde{K}_{22}^{\mathrm{non}}$ as functions of $S$. The ratio $\tilde{K}_{33}^{\mathrm{non}}/\tilde{K}_{22}^{\mathrm{non}}$ increases rapidly with increasing orientational order, demonstrating that the bend elasticity becomes more dominant in the highly ordered nematic state. In contrast, we have the ratio $\tilde{K}_{11}^{\mathrm{non}}/\tilde{K}_{22}^{\mathrm{non}}=3$ for the entire range of the order parameter. 
The filled circles in Fig. \ref{fig2}(b) represent the numerical results reported by Lee and Meyer~\cite{Lee85}. The present calculations successfully reproduce these results.
At $\phi\simeq 0.3$ in Fig.~\ref{fig2}(a), we have $K_{33}^{non}/c_1\simeq 0.017$ and $S=0.825$. Then, for the parameter set $D=25$ $\AA$, $T=300$ K, and $n=L/D=20$, we estimate $K_{33}^{non}\simeq 10.74~{\rm pN}$, which is of the same order of magnitude as the Straley model~\cite{Straley73}.
Consistent with the Straley theory, we obtain the relation $K_{33}^{non}>K_{11}^{non}>K_{22}^{non}$ and $K_{11}^{non}/K_{22}^{non}=3$ throughout the concentration range, which are characteristic features of the Straley model~\cite{Straley73,Lee85,Odijk86}.

\subsection{Monomer-dimer model}
\label{sub2}
We here consider the simplest associating system, consisting of monomers and dimers, to clarify how aggregation affects the Frank elastic constants. In this minimal aggregation model, only monomeric and dimeric species ($m,l=1,2$) are taken into account. Despite its simplicity, the model captures the essential excluded-volume interaction between aggregates of different lengths.

The total volume fraction $\phi$ of the solute molecules in the monomer-dimer model is given by
\begin{equation}
\phi=\phi_1+\phi_2.
\label{md1}
\end{equation}
Using Eq. (\ref{eq220c}),  the volume fraction $\phi_1$ of monomers is given by
\begin{equation}
\phi_1=\frac{1}{4K_a }\big[-1+\sqrt{1+8\phi K_a}\big],
\label{md2}
\end{equation}
and the mean aggregation number (Eq. (\ref{eq217a})) becomes
\begin{equation}
\langle m \rangle=\phi/(\phi_1+\phi_2/2).
 \label{eq217aaa}
\end{equation}
For the present monomer--dimer numerical calculations, we take $K_a=100$, $n=L_1/D=20$ for a monomer, and $L_2/D=40$ for a dimer.

\begin{figure}
\begin{center}
\includegraphics*[height=5cm]{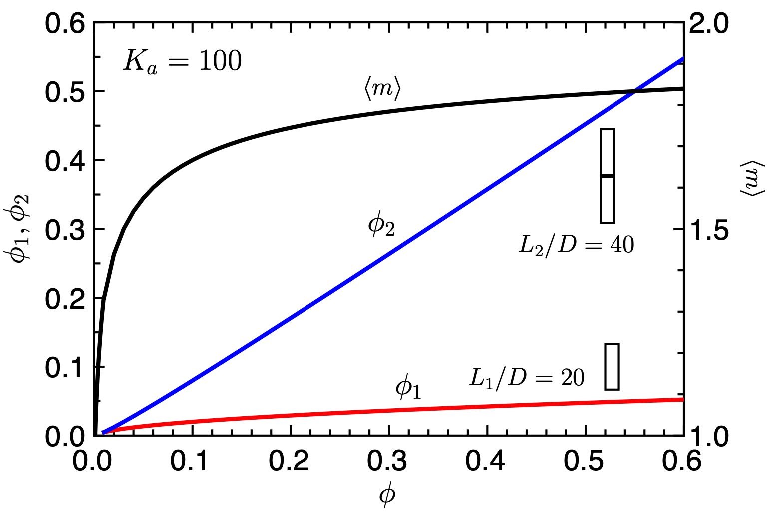}
\caption{(Colour online) Concentration dependence of the monomer and dimer volume fractions, \(\phi_1\) and \(\phi_2\), together with the average aggregation number \(\langle m \rangle\) in the monomer-dimer model.
}
\label{fig3}
\end{center}
\end{figure}

Figure~\ref{fig3} shows the concentration dependence of the monomer volume fraction $\phi_1$, the dimer volume fraction $\phi_2$, and the mean aggregation number $\langle m \rangle$. As the total concentration $\phi$ increases, the dimer fraction grows rapidly, whereas the monomer fraction remains relatively small over the entire concentration range. The mean aggregation number increases from $\langle m \rangle =1$ toward the dimer limit $\langle m \rangle =2$.
These results indicate that the system becomes dominated by dimers at high concentrations. 

\begin{figure}
\begin{center}
\includegraphics*[height=5cm]{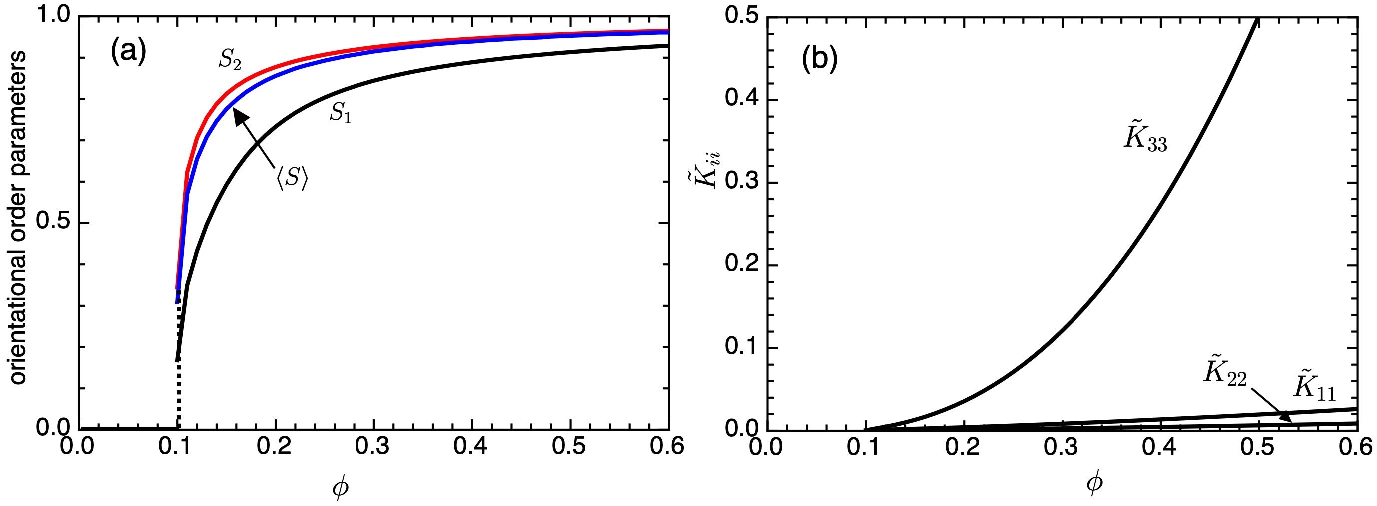}
\caption{(Colour online) Orientational order parameters $S_1$, $S_2$, and the averaged one \(\langle S \rangle\) (a) and the reduced Frank elastic constants $\tilde{K}_{ii}=K_{ii}/c_1$ (b), plotted against the volume fraction $\phi$ in the monomer-dimer model.
}
\label{fig4}
\end{center}
\end{figure}

Figure~\ref{fig4}(a) shows the concentration dependence of the orientational order parameters for monomers $S_1$, dimers $S_2$, and the average orientational order parameter $\langle S \rangle$. At the NIT concentration $\phi\simeq 0.12$, both monomers and dimers exhibit discontinuous increases in orientational order. The dimer order parameter $S_2$ is consistently larger than the monomer order parameter $S_1$. The average orientational order parameter lies between these two limits and increases with concentration.
The corresponding reduced Frank elastic constants $\tilde{K}_{ii}=K_{ii}/c_1$ are shown in Fig.~\ref{fig4}(b). The bend elastic constant $\tilde{K}_{33}$ increases much more steeply with concentration than the splay and twist elastic constants and becomes the dominant elastic mode in the nematic phase. 
These results indicate that even the simplest aggregation process strongly enhances the bend elasticity. Since dimers possess larger excluded volumes than monomers, the orientational coupling between dimers and surrounding molecules becomes increasingly important with increasing concentration. The presence of longer aggregates enhances the collective orientational coherence of the system, particularly for the bend mode, where longitudinal orientational contributions survive under strong alignment conditions.

\begin{figure}
\begin{center}
\includegraphics*[height=5cm]{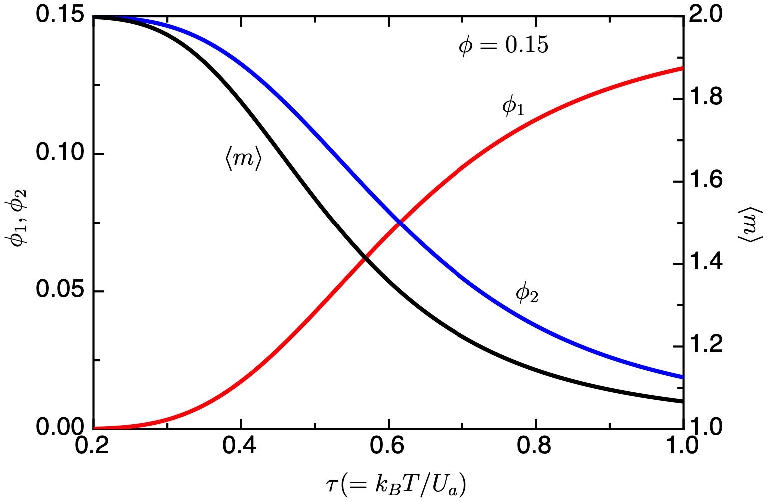}
\caption{(Colour online) Monomer volume fraction $\phi_1$, the dimer volume fraction $\phi_2$, and the mean aggregation number $\langle m \rangle$, plotted against the reduced temperature $\tau$ at the fixed concentration $\phi=0.15$ in the monomer-dimer model.
}
\label{fig5}
\end{center}
\end{figure}

We next consider the temperature dependence of the orientational order and the Frank elastic constants in the monomer--dimer system. In the present model, the aggregation equilibrium is strongly affected by temperature through the association constant $K_a$. For the numerical calculations, we introduce the dimensionless reduced temperature
$
\tau={k_BT}/{U_a},
$
where $U_a$ denotes a characteristic interaction energy.
We further define the bonding energy parameter
$
\gamma={\epsilon}/{U_a},
$
so that the association constant (Eq.~(\ref{eq220dd})) can be written as
\begin{equation}
K_a=\omega \exp(\gamma/\tau),
\label{md1a}
\end{equation}
which is a function of the reduced temperature $\tau$. In the following calculations, we set $\omega=0.01$, $\gamma=5$, and $\phi=0.15$ as representative parameter values, used in the previous paper~\cite{Matsu25}.

\begin{figure}
\begin{center}
\includegraphics*[height=5cm]{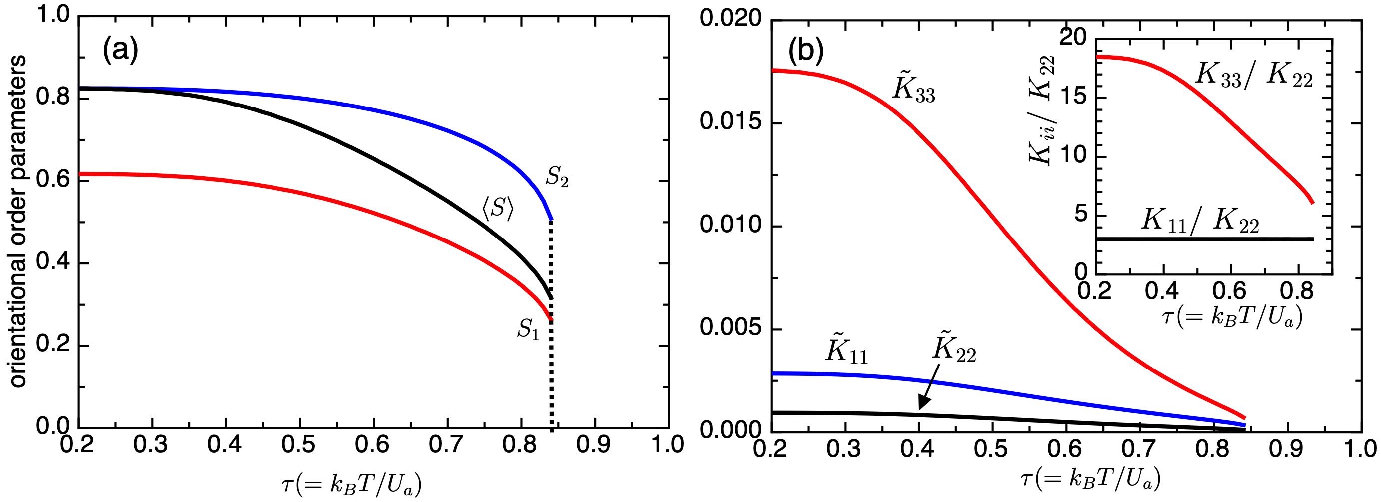}
\caption{(Colour online) Orientational order parameters $S_1$, $S_2$, and $\langle S \rangle$ (a), and the reduced Frank elastic constants $\tilde{K}_{ii}$ (b), plotted as functions of the reduced temperature $\tau$ at $\phi=0.15$ in the monomer-dimer model.
}
\label{fig6}
\end{center}
\end{figure}

Figure~\ref{fig5} shows the temperature dependence of the monomer volume fraction $\phi_1$, the dimer volume fraction $\phi_2$, and the mean aggregation number $\langle m \rangle$. As the reduced temperature $\tau$ decreases, the dimer fraction increases while the monomer fraction decreases, indicating that aggregation becomes progressively more favorable at lower temperatures. The mean aggregation number increases from $\langle m \rangle \simeq 1$ at high temperatures toward the dimer limit $\langle m \rangle =2$.
These results demonstrate that decreasing temperature promotes aggregation by increasing the association constant $K_a$.

Figure~\ref{fig6}(a) shows the temperature dependence of the orientational order parameters $S_1$, $S_2$, and the average orientational order parameter $\langle S \rangle$. 
The system undergoes a first-order NIT at $\tau \simeq 0.84$, where the order parameters exhibit discontinuous jumps.
As the reduced temperature $\tau$ decreases, all order parameters increase monotonically due to the enhanced orientational ordering induced by aggregation. The corresponding reduced Frank elastic constants $\tilde K_{ii}$ are shown in Fig.~\ref{fig6}(b). All elastic constants increase with decreasing temperature. The bend elastic constant $\tilde K_{33}$ exhibits a much stronger temperature dependence than the splay and twist elastic constants and becomes the dominant elastic mode in the low-temperature nematic phase.
The inset of Fig.~\ref{fig6}(b) shows the elastic anisotropy ratios. While the ratio $K_{11}/K_{22}$ remains constant at the Straley value $3$, the ratio $K_{33}/K_{22}$ increases with decreasing temperature. 
These results demonstrate that aggregation selectively enhances bend elasticity by increasing the dimer population.
In the low-temperature limit, $\tilde{K}_{ii}$ is expected to approach the values for a nonassociating rod system with $L/D=40$, 
since the system becomes dominated by dimers as shown in Fig.~\ref{fig5}.

The present results demonstrate that dimer formation provides an efficient mechanism for controlling the elastic anisotropy of nematic liquid crystals. In particular, the bend elastic constant is selectively and substantially enhanced, whereas the splay and twist elastic constants exhibit only moderate changes.

\subsection{Rodlike aggregates model}
\label{sub3}

In this subsection, we present numerical results for the rodlike aggregates model, in which all aggregation states ($l,m=1,2,\cdots,\infty$) are taken into account. We set the monomer aspect ratio to $n=L/D=1$. Using Eqs.~(\ref{eq203}) and (\ref{eq220c}), the monomer volume fraction is obtained as
\begin{equation}
\phi_1=\frac{1}{2K_a^2\phi}\bigg(1+2K_a\phi-\sqrt{1+4K_a\phi}\bigg),
\label{eq_agg1}
\end{equation}
while the mean aggregation number, defined by Eq.~(\ref{eq217a}), is given by
\begin{equation}
\langle m \rangle=\frac12\big(1+\sqrt{1+4K_a\phi}\big).
\label{eq_agg2}
\end{equation}

\begin{figure}
\begin{center}
\includegraphics*[height=5cm]{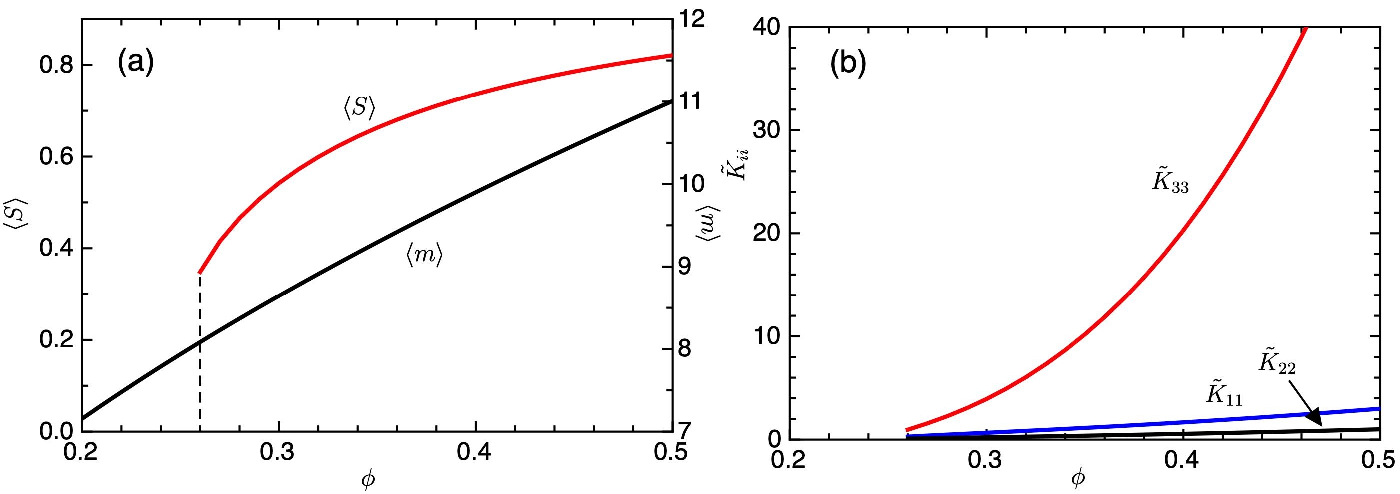}
\caption{(Colour online) Concentration dependence of the average orientational order parameter $\langle S \rangle$, the mean aggregation number $\langle m \rangle$ (a), and the reduced elastic constants ${\tilde K}_{ii}$ (b) at the reduced temperature $\tau=0.5$.
}
\label{fig7}
\end{center}
\end{figure}

We first show the concentration dependence of the Frank elastic constants for the rodlike aggregates model with $\omega=0.01$ and $\gamma=5$ in Eq.~(\ref{md1a}). 
Figure~\ref{fig7}(a) shows the concentration dependence of the average orientational order parameter $\langle S \rangle$ and the mean aggregation number $\langle m \rangle$ at the reduced temperature $\tau=0.5$. The system undergoes the NIT at $\phi \simeq 0.26$, above which the value of $\langle S \rangle$ increases with concentration. 
The mean aggregation number $\langle m \rangle$ increases with concentration, indicating the formation of long rodlike aggregates.

Figure~\ref{fig7} (b) shows the concentration dependence of the reduced elastic constants ${\tilde K}_{ii}$.
All elastic constants increase with concentration in the nematic phase. The bend elastic constant $\tilde K_{33}$ exhibits rapid growth and becomes much larger than the splay and twist elastic constants at high concentrations.
These results indicate that long aggregates strongly enhance the bend elasticity through cooperative excluded-volume interactions. Since the excluded-volume interaction scales approximately as $L_mL_l$, long aggregates dominate the bend elasticity at high concentrations.

The large values of the bend elastic constant obtained in the present calculations originate from the rigid-rod aggregate limit adopted in the model. Because the aggregates are assumed to behave as perfectly rigid rods, long aggregates make large contributions to the excluded-volume interaction, resulting in a large enhancement of the bend elasticity.
For example, when $D=25~\AA$, $T=300$ K, $n=1$, and $K_{33}/c_1=40$ at $\phi\simeq 0.46$ in Fig.~\ref{fig7}(b), we obtain
$
K_{33}\simeq 107.6\ {\rm pN},
$
which is considerably larger than typical values observed in conventional nematic liquid crystals. Giant bend elastic constants in the range of $10^2$--$10^3$ pN have also been experimentally reported for nematic liquid crystals containing cybotactic smectic clusters~\cite{Aya14}.
Self-assembled rodlike aggregates have also been experimentally observed in short DNA duplex systems, where end-to-end stacking gives rise to nematic and columnar liquid-crystalline phases~\cite{Nakata07}. The existence of liquid-crystalline ordering in such self-assembled systems demonstrates that sufficiently rigid aggregates can indeed generate strong orientational correlations. The present results should therefore be regarded as upper-bound estimates, since the theory assumes perfectly rigid aggregates and neglects aggregate flexibility, finite persistence lengths, and other effects that would reduce the elastic response.

\begin{figure}
\begin{center}
\includegraphics*[height=5cm]{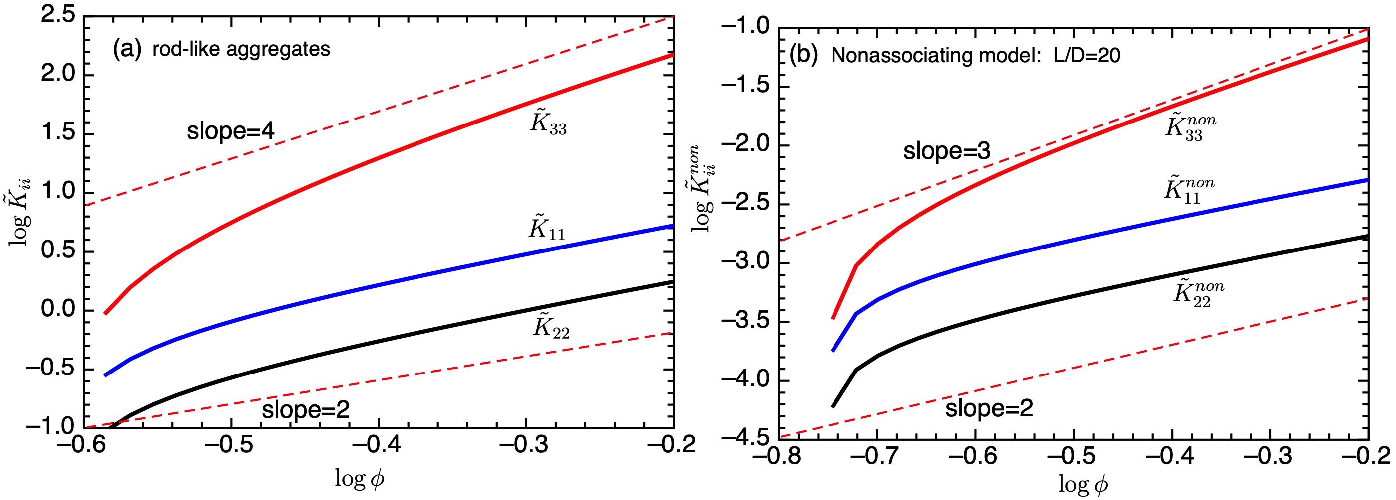}
\caption{(Colour online) Log-log plots of the concentration dependence of the reduced Frank elastic constants for (a) the rodlike aggregates model and (b) the nonassociating Straley model with $L/D=20$.
}
\label{fig8}
\end{center}
\end{figure}
Figure~\ref{fig8} shows log--log plots of the concentration dependence of the reduced Frank elastic constants for (a) the rodlike aggregates model of Fig.~\ref{fig7}(b) and (b) the nonassociating model with $L/D=20$ of Fig.~\ref{fig2}(a). In the rodlike aggregates model, the splay and twist elastic constants exhibit approximately quadratic concentration dependence,
$
K_{11},K_{22}\propto\phi^2,
$
whereas the bend elastic constant shows a much stronger concentration dependence and approaches
$
K_{33}\propto\phi^4
$
at higher concentrations. In contrast, the numerical results for the nonassociating model show approximately quadratic concentration dependence of $K_{11}^{\rm non}$ and $K_{22}^{\rm non}$, and approximately cubic concentration dependence of $K_{33}^{\rm non}$ over the concentration range considered. The latter behavior is consistent with the strong-alignment scaling predicted by Odijk~\cite{Odijk86}. Thus, aggregation changes the concentration dependence of the bend elastic constant, while the splay and twist elastic constants remain approximately quadratic in concentration.
In both systems, the relation $K_{11}=3K_{22}$ holds throughout the entire concentration range. This relation is a consequence of the present Straley-type treatment, which accounts solely for excluded-volume interactions. Additional contributions to the splay elastic constant arising from the entropic effects of chain ends and density fluctuations, as discussed in theories of polymer liquid crystals~\cite{Meyer82}, are not included in the present model.

The origin of the different scaling behaviors in Fig.~\ref{fig8}(a) can be understood from the structure of Eq.~(\ref{eq309}):
\begin{equation}
K_{ii}\sim\sum_{m,l}\phi_m\phi_l
\iint f_m'f_l'\tilde G_i(m,l)d\Omega d\Omega',
\end{equation}
where $\tilde G_i(m,l)$ (Eq.~(\ref{eq3010})) denotes the orientational kernel for the splay, twist, and bend modes. For the splay and twist modes, the kernels $\tilde G_1$ and $\tilde G_2$ mainly contain transverse orientational components associated with $\Omega_y$ and $\Omega_z$. Under strong alignment conditions, these transverse components become small, and the corresponding angular integrals saturate, exhibiting only a weak dependence on concentration. The concentration dependence originates primarily from the prefactor $\phi_m\phi_l$, leading to the approximate scaling behavior $K_{11}, K_{22}\propto\phi^2$. In contrast, the bend kernel $\tilde G_3$ contains 
orientational contributions along the alignment axis (the $x$-axis) associated with $\Omega_x\simeq1$, which survive even in the strong-alignment limit. Moreover, the derivatives of the orientational distribution functions satisfy approximately $f_m'f_l'\propto ml\phi^2$, while $\tilde G_3$ contains factors proportional to $m^2$ and $l^2$. As a result, higher-order moments of the volume fraction distribution contribute strongly to the bend elasticity, weighting long aggregates much more heavily than short ones. Because aggregation shifts the volume fraction distribution $\phi_m$ toward larger $m$ with increasing total concentration $\phi$, it strongly amplifies the contribution of the long-tail part of the distribution, yielding the steeper concentration dependence approaching $K_{33}\propto\phi^4$. This giant bend elasticity can thus be attributed to the combined effects of longitudinal orientational contributions and cooperative excluded-volume interactions associated with long aggregates.

The concentration dependence of the nonassociating model shown in Fig.~\ref{fig8}(b) can likewise be understood from the angular integral,
$
\iint f'f'\tilde G_i\,d\Omega d\Omega',
$
corresponding to the monodisperse case $m=l=1$. As concentration increases, the orientational distribution function becomes increasingly sharp. For the splay and twist modes, the transverse components scale approximately as $\Omega_y,\Omega_z\sim\theta$, so that the resulting angular integrals remain small and nearly insensitive to $\phi$ in the concentration regime considered here. Thus, the prefactor $\phi^2$ dominates, giving the approximately quadratic behavior observed in Fig.~\ref{fig8}(b). Note that while Odijk's asymptotic theory predicts $K_{11}, K_{22}\propto\phi$ in the extreme strong-alignment limit where the angular integral decays as $\phi^{-1}$~\cite{Odijk86}, the present numerical evaluation spans an intermediate-to-high concentration regime (up to $\phi \simeq 0.6$), where the effective exponent remains closer to $\phi^2$. In contrast, for the bend mode, although the longitudinal contribution ($\Omega_x\simeq1$) is not enhanced by aggregate growth since $m=l=1$, the sharpening of the orientational distribution causes the angular integral itself to increase approximately linearly with concentration:
$
\iint f'f'\tilde G_3\,d\Omega d\Omega'\propto\phi.
$
Combined with the excluded-volume prefactor $\phi^2$, this yields $K_{33}^{\rm non}\propto\phi^3$, in agreement with Odijk's strong-alignment scaling~\cite{Odijk86}.

\begin{figure}[h]
\begin{center}
\includegraphics*[height=10.cm]{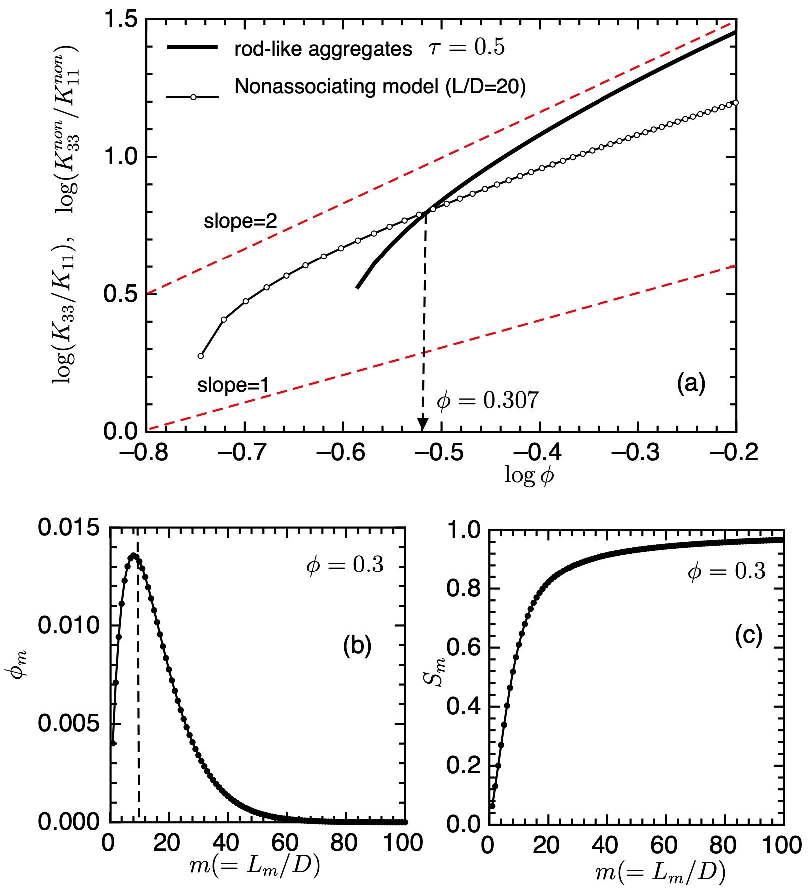}
\caption{(Colour online) (a) Comparison of the elastic anisotropy ratio $K_{33}/K_{11}$ between the rodlike aggregates model and the nonassociating model with $L/D=20$. (b) Volume fraction distribution of aggregates, $\phi_m$ and (c) orientational order parameter, $S_m$, at $\phi=0.3$ and $\tau=0.5$. The long-tail structure of the volume fraction distribution and the strong alignment of long aggregates play important roles in the enhancement of the bend elasticity.
}
\label{fig9}
\end{center}
\end{figure}

Figure~\ref{fig9}(a) compares the elastic anisotropy ratio $K_{33}/K_{11}$ for the rodlike aggregates model at $\tau=0.5$ with $K_{33}^{\mathrm{non}}/K_{11}^{\mathrm{non}}$ for the nonassociating model with $L/D=20$, shown in Fig.~\ref{fig8}. In the nonassociating system, the ratio increases linearly with concentration at high concentrations. In contrast, the aggregating system exhibits a much steeper increase, with approximately $K_{33}/K_{11}\propto\phi^2$ at high concentrations. The two ratios become equal at a crossover concentration of $\phi\simeq0.307$, where both ratios are approximately 6. Above this concentration, the elastic anisotropy of the aggregating system exceeds that of the nonassociating model.
The comparison shows that, although the absolute values of the Frank elastic constants differ between the two systems, their elastic anisotropy ratios can become identical. This indicates that aggregation can produce an elastic anisotropy comparable to that of intrinsically elongated rods, even when the constituent monomers are relatively short.

Figures~\ref{fig9}(b) and (c) show the volume fraction distribution $\phi_m$ and the orientational order parameter $S_m$, respectively, at the crossover concentration $\phi=0.3$. The size distribution in Fig.~\ref{fig9}(b) exhibits a peak around $m\simeq9$ together with a pronounced long tail extending to larger aggregation numbers. The orientational order parameter $S_m$ in Fig.~\ref{fig9}(c) increases with $m$. As shown in Fig.~\ref{fig7}(a), the mean aggregation number at this concentration is only $\langle m\rangle\simeq 9$. 
The long-tail region with $m>9$ makes a large contribution to the bend elasticity because aggregates in this region combine strong excluded-volume interactions with high orientational order. These results suggest that the giant bend elasticity of associating rodlike systems is governed not only by the average aggregation number itself but by the long-tail structure of the volume fraction distribution.
In the case of a monodisperse system with $n=9$ and $l, m=1$, all rods have the same length as the mean aggregate length in the associating system. In this case, the monodisperse system remains isotropic, with $S=0$, whereas the associating system exhibits nematic order. 
This comparison indicates that aggregate-size polydispersity plays an important role in promoting nematic ordering, even when the mean aggregation number is relatively small.

\begin{figure}
\begin{center}
\includegraphics*[height=6.cm]{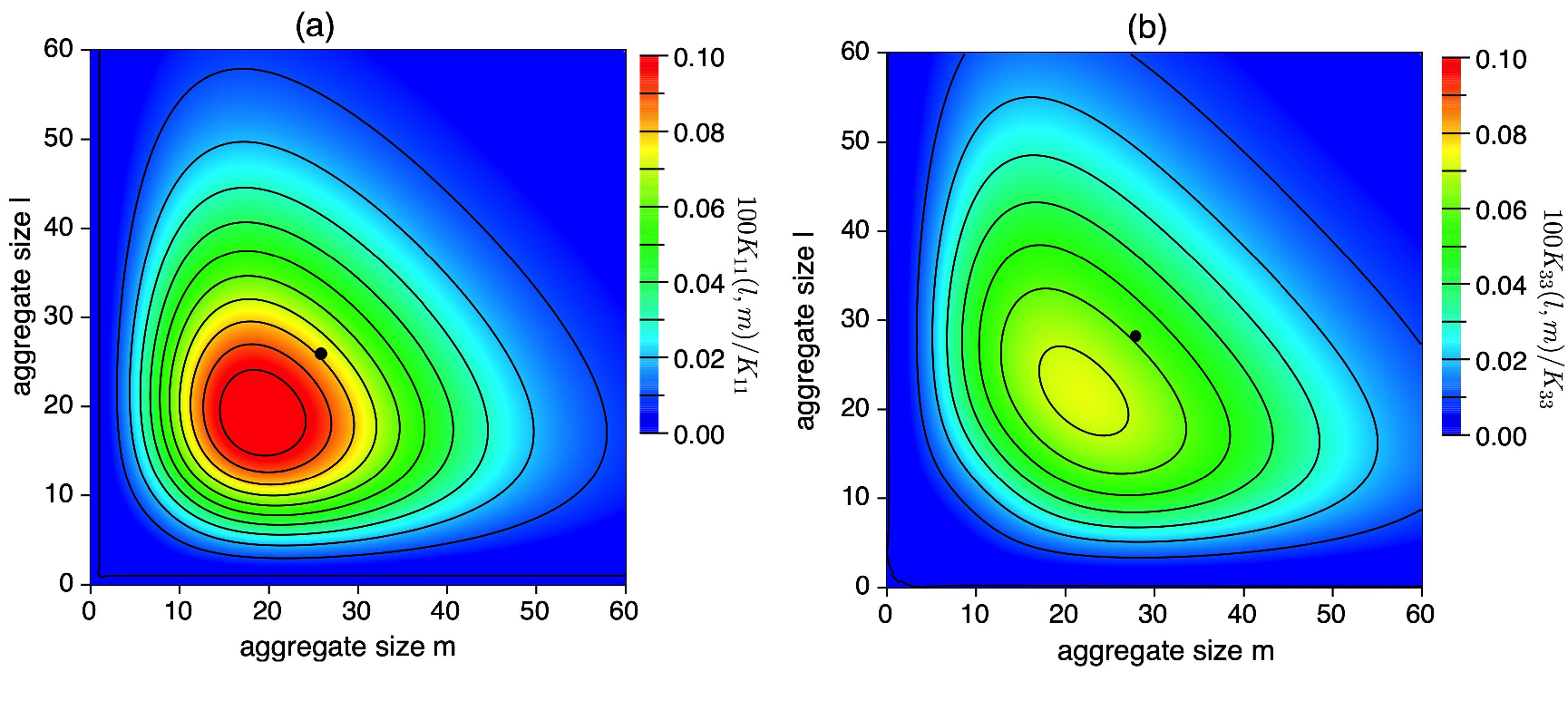}
\caption{(Colour online) Contour plots of the normalized $l-m$ pair contributions to the Frank elastic constants at $\tau=0.5$ and $\phi=0.3$ in Fig.~\ref{fig7}(b): (a) splay elastic constant $K_{11}$ and (b) bend elastic constant $K_{33}$. The color scale represents $100K_{ii}(l,m)/K_{ii}$, where $K_{ii}(l,m)$ is the contribution from a pair of aggregates of sizes $l$ and $m$, and $K_{ii}$ is the corresponding total elastic constant. The black circles indicate the contribution-weighted centroids of the respective distributions.
}
\label{fig10}
\end{center}
\end{figure}

Figure~\ref{fig10} shows the $l-m$ pair-resolved distributions of the splay and
bend elastic constants at $\phi=0.3$ in Figs.~\ref{fig7} and~\ref{fig9} :
(a) $K_{11}(l,m)$ and (b) $K_{33}(l,m)$.
The color scale represents the normalized contribution $100K_{ii}(l,m)/K_{ii}$, expressed as a percentage, from pairs of aggregates with sizes
$l$ and $m$. 
The pair-resolved contributions $K_{ii}(l,m)$ are proportional
to the product $\phi_l\phi_m$ multiplied by the corresponding the geometrical kernel ${\tilde G}_{i}(l,m)$, which includes terms proportional to
$l^2$ and $m^2$, shown in Eqs.~(\ref{eq311})--(\ref{eq313}). Thus, the size dependence of the
pair contributions is determined by both the aggregate volume fraction
distribution and the size dependence of the kernel.
The average aggregation number under these
conditions is $\langle m\rangle\simeq 9$, as shown in Fig.~\ref{fig9}(b).
Nevertheless, the contributions to both elastic constants
extend to larger aggregate sizes, reflecting
the long-tailed volume fraction distribution.
We find a clear difference between the two distributions.
The contribution to the splay elastic constant $K_{11}$ is
relatively localized around $l=m\simeq 20$, whereas that to the bend elastic constant
$K_{33}$ extends over a broader range of aggregate sizes.
For example, for a fixed aggregate size of $m=10$, the
contribution to $K_{11}$ extends up to approximately $l\sim 50$,
while that to $K_{33}$ remains appreciable up to approximately
$l\sim 60$. This comparison indicates that the contribution to
the bend elastic constant extends further into the long-tail
region of the volume fraction distribution than that to the
splay elastic constant.
The black circles in Fig.~\ref{fig10} indicate the contribution-weighted centroids of the respective distributions, defined by
\begin{equation}
\langle m\rangle_{K_{ii}}
=
\frac{\displaystyle\sum_{l,m} mK_{ii}(l,m)}
{\displaystyle\sum_{l,m}K_{ii}(l,m)}.
\label{eq:pair_centroid}
\end{equation}
Since the $l-m$ pair contributions are symmetric with respect to the interchange of $l$ and $m$, the two centroid coordinates are identical. The mean aggregate size characterizing the $K_{33}$ contribution is $\langle m\rangle_{K_{33}}=29.44$, which is 12.7\% larger than that characterizing the $K_{11}$ contribution: $\langle m\rangle_{K_{11}}=26.13$. This indicates that the distribution of $K_{33}$ is shifted toward larger aggregate sizes compared with that of $K_{11}$. We also find that the pair-resolved distribution of the twist elastic constant $K_{22}(l,m)$ is almost identical to that of the splay elastic constant $K_{11}(l,m)$. Therefore, the distribution of $K_{22}$ is not shown. The contribution-weighted centroid of $K_{22}$ is $\langle m\rangle_{K_{22}}=25.71$, which is close to that of $K_{11}$.
The calculations are performed for aggregate sizes up to
$l,m=100$, which capture approximately 99\% of the total
value of each Frank elastic constant. The aggregate-size
distribution is truncated when the relative population
fraction satisfys $\phi_m/\phi < 10^{-12}$. The contour plots of Fig.~\ref{fig10} are displayed over the range $ l,m\leq60$.

We finally show the temperature dependence of the Frank elastic constants in the rodlike aggregates system. 
The aggregation equilibrium is controlled by temperature through the association constant $K_a$ (Eq.~(\ref{md1a})) with $\omega=0.01$ and $\gamma=5$.

\begin{figure}
\begin{center}
\includegraphics*[height=5.cm]{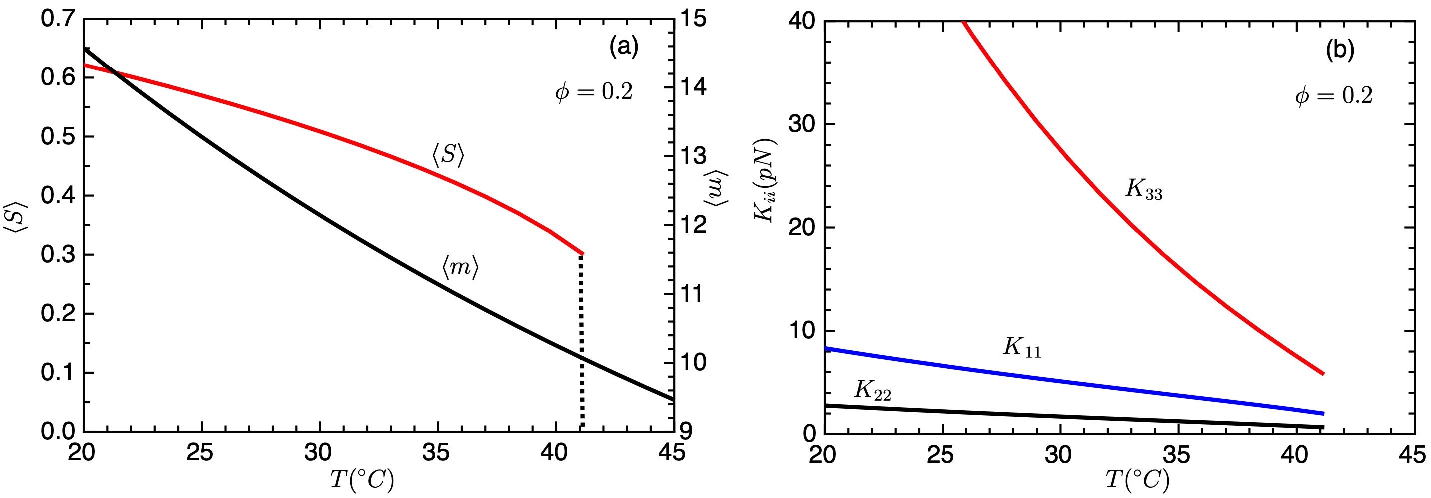}
\caption{(Colour online) (a) Temperature dependence of the average orientational order parameter $\langle S\rangle$, the mean aggregation number $\langle m\rangle$, and (b) the corresponding temperature dependence of the dimensional elastic constants $ K_{ii}$ (pN), at the fixed concentration $\phi=0.2$.
}
\label{fig11}
\end{center}
\end{figure}
Figure~\ref{fig11} shows the temperature dependence of
$\langle S\rangle$ and $\langle m\rangle$ (a), together
with the dimensional Frank elastic constants $K_{ii}$
in pN (b), at the fixed concentration $\phi=0.2$.
The temperature scale is obtained by converting the
reduced temperature $\tau$ using $U_a/k_B=674$ K and the NIT temperature $T_{NI}=41^{\circ}C$.
As the temperature decreases, the association constant
increases, leading to an increase in the mean aggregation
number, as shown in Fig.~\ref{fig11}(a). All three elastic constants decrease with
increasing temperature, as shown in Fig.~\ref{fig11}(b). The bend elastic constant $K_{33}$ exhibits a much stronger temperature
dependence than the splay and twist elastic constants,
while the relation $K_{11}=3K_{22}$ is preserved throughout
the entire temperature range.
This behavior reflects the fact that long aggregates contribute strongly to the longitudinal orientational correlations through excluded-volume interactions, thereby leading to an enhancement of the bend elastic constant. Wensink~\cite{Wensink19} showed that increasing the persistence length of associating rods leads to an anomalous increase in the bend elastic modulus, which becomes the principal modulus in the rigid-rod limit.

It should be noted that Fig.~\ref{fig6}(b) shows the temperature dependence of the reduced elastic constants
$\widetilde{K}_{ii}=K_{ii}/c_1$, whereas
Fig.~\ref{fig11}(b) shows that of the dimensional elastic
constants $K_{ii}$. Although the conversion factor
$c_1(T)$ increases linearly with temperature, the decrease
in the reduced elastic constants $\widetilde{K}_{ii}$ with
increasing temperature is sufficiently large, so that the
dimensional Frank elastic constants $K_{ii}$ also
decrease as the temperature approaches $T_{NI}$.

Experiments on self-assembled chromonic systems~\cite{Zhou14,Zhou12} have reported
$K_{11}\simeq K_{33}$, in contrast to conventional
rodlike nematics. This difference suggests that
aggregate flexibility, finite persistence lengths,
and contour fluctuations~\cite{Vroege88,Ferrarini18},
which are neglected in the present theory, may
enhance the splay elasticity and suppress the bend
elasticity. The present results should therefore be
regarded as representing an upper-bound elastic
response in the rigid-aggregate limit.

\section{Summary}
\label{sec4}
In this paper, we have theoretically studied the Frank elastic constants of associating rodlike systems using a mean field theory combined with the Straley elastic free energy formalism. 
By incorporating the equilibrium aggregate volume fraction distribution with a self-consistent treatment of the size-dependent orientational ordering, we have systematically analyzed the effects of aggregation on the elastic properties for both a monomer--dimer model and a fully polydisperse rodlike aggregate model.

For the monomer--dimer model, we show that even the simplest aggregation process enhances the bend elastic constant. As the dimer population increases, aggregation strengthens the orientational ordering and selectively amplifies the bend elasticity relative to the splay and twist elasticities.

For the rodlike aggregate model, we find an enhancement of the bend elastic constant induced by aggregation, resulting in the elastic anisotropy $K_{33}\gg K_{11}>K_{22}$. Owing to the long-tail structure of the volume fraction distribution of the aggregates, a small population of long aggregates can make large contributions to the excluded-volume interactions. This effect leads to a strong enhancement of the collective orientational response, particularly for bend deformation. 
The elastic constants also exhibit approximate scaling behaviors at higher concentrations, with $K_{11}, K_{22}\propto\phi^{2}$ and
$K_{33}\propto\phi^{4}$.
We also find that decreasing temperature promotes the formation of long rodlike aggregates and induces a rapid increase in the bend elastic constant, while the relation $K_{11}=3K_{22}$ remains unchanged irrespective of temperature and concentration.

Our theory is based on the limiting case of perfectly rigid rodlike aggregates undergoing ideal linear aggregation. In associating liquid-crystalline systems, however, these assemblies generally possess finite persistence lengths and internal flexibility~\cite{Wensink19,Doussal91}. For example, chromonic liquid crystals and self-assembled DNA duplex systems form rodlike aggregates whose elastic properties can be affected by contour fluctuations and aggregate flexibility~\cite{Vroege88,Ferrarini10,Ferrarini18}. 
Because these factors are neglected in the present model, our predictions represent the rigid-rod limit of the bend elastic constant in associating rodlike systems.


\end{document}